\documentclass{article}
\usepackage{amsfonts}
\usepackage{spconf,amsmath,graphicx,epsfig}
\usepackage{subcaption}
\usepackage{mwe}
\usepackage{xcolor}
\usepackage{hyperref}
\usepackage{algorithm}
\usepackage{algorithmic}
\usepackage{multirow}
\usepackage{braket}
\usepackage{qcircuit}
\usepackage{adjustbox}

\usepackage{color}

\title{Decentralizing Feature Extraction with Quantum Convolutional Neural Network for Automatic Speech Recognition}

\name{%
\begin{tabular}{@{}c@{}}
Chao-Han Huck Yang$^{1}$\qquad Jun Qi$^{1}$\qquad Samuel Yen-Chi Chen$^{2}$\qquad Pin-Yu Chen$^{3}$ \\ Sabato Marco Siniscalchi$^{1,4,5}$\qquad Xiaoli Ma$^{1}$\qquad Chin-Hui Lee$^{1}$
\end{tabular}}
\address{$^1$School of Electrical and Computer Engineering, Georgia Institute of Technology, USA\\$^2$Brookhaven National Laboratory, NY, USA and $^3$IBM Research, Yorktown Heights, NY, USA \\ $^4$Faculty of Computer and Telecommunication Engineering, University of Enna, Italy \\ $^5$Department of Electronic Systems, NTNU, Trondheim, Norway}

\begin{document}
\ninept
\maketitle
\begin{abstract}
We propose a novel decentralized feature extraction approach in federated learning to address privacy-preservation issues for speech recognition. It is built upon a quantum convolutional neural network (QCNN) composed of a quantum circuit encoder for feature extraction, and a recurrent neural network (RNN) based end-to-end acoustic model (AM). To enhance model parameter protection in a decentralized architecture, an input speech is first up-streamed to a quantum computing server to extract Mel-spectrogram, and the corresponding convolutional features are encoded using a quantum circuit algorithm with random parameters. The encoded features are then down-streamed to the local RNN model for the final recognition. The proposed decentralized framework takes advantage of the quantum learning progress to secure models and to avoid privacy leakage attacks. Testing on the Google Speech Commands Dataset, the proposed QCNN encoder attains a competitive accuracy of 95.12\% in a decentralized model, which is better than the previous architectures using centralized RNN models with convolutional features. We conduct an in-depth study of different quantum circuit encoder architectures to provide insights into designing QCNN-based feature extractors. Neural saliency analyses demonstrate a high correlation between the proposed QCNN features, class activation maps, and the input Mel-spectrogram. We provide an implementation\footnote{\href{https://github.com/huckiyang/QuantumSpeech-QCNN}{https://github.com/huckiyang/QuantumSpeech-QCNN}} for future studies.

\end{abstract}

\section{Introduction}
\label{sec1}
With the increasing concern about acoustic data privacy issues~\cite{leroy2019federated}, it is essential to design new automatic speech recognition (ASR) architectures satisfying the requirements of new privacy-preservation regulations, e.g., GDPR~\cite{voigt2017eu}. Vertical federated learning (VFL)~\cite{yang2019federated} is one potential strategy for data protection by decentralizing an end-to-end deep learning~\cite{deng2013new} framework and separating feature extraction from the ASR inference engine. With recent advances in commercial quantum technology~\cite{mohseni2017commercialize}, quantum machine learning (QML)~\cite{mitarai2018quantum} becomes an ideal building block for VFL owing to its advantages on parameter encryption and isolation. To do so, the input to QML often represented by classical bits, needs to be first encoded into quantum states based on \emph{qubits}. Next, approximation algorithms (e.g., quantum branching programs~\cite{bergholm2018pennylane}) are applied to quantum devices based on a quantum circuit~\cite{havlivcek2019supervised} with noise tolerance. To implement our proposed approach, we utilize a state-of-the-art noisy intermediate-scale quantum (NISQ)~\cite{farhi2000quantum} platform (5 to 50 qubits) for academic and commercial applications~\cite{preskill2018quantum}. It can be set up on accessible quantum servers from cloud-based computing providers~\cite{mohseni2017commercialize}.

\begin{figure}[t]
\begin{center}
\vspace{-2mm}
  \includegraphics[width=0.90\linewidth]{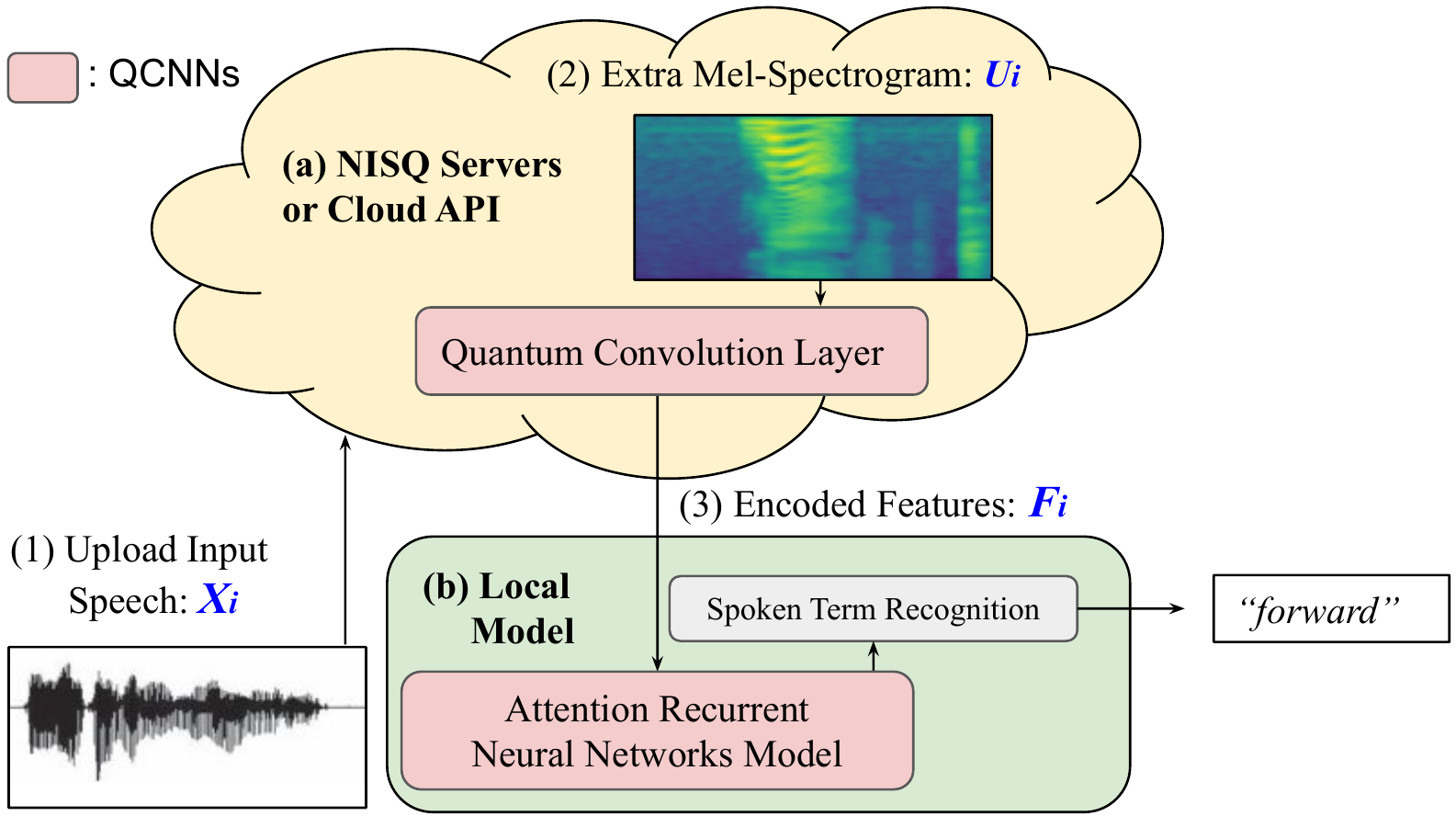}
\end{center}
  \caption{Proposed quantum machine learning for acoustic modeling (QML-AM) architecture in a vertical federated learning progress including (a) a quantum convolution layer on Noisy Intermediate-Scale Quantum (NISQ) servers or cloud API; and (b) a local model (e.g., second-pass model~\cite{ chen2019federated,yang2020multi}) for speech recognition tasks. 
  } 
\label{fig:1:vfl}
\end{figure}

As shown in Fig.~\ref{fig:1:vfl}, we propose a decentralized acoustic modeling (AM) scheme to design a quantum convolutional neural network (QCNN)~\cite{henderson2020quanvolutional} by combining a variational quantum circuit (VQC) learning paradigm~\cite{mitarai2018quantum} and a deep neural network~\cite{hochreiter1997long} (DNN). VQC refers to a quantum algorithm with a flexible designing accessibility, which is resistant to noise~\cite{mitarai2018quantum,havlivcek2019supervised} and adapted to NISQ hardware with light or no requirements for quantum error correction. Based on the advantages of VQC under VFL, a quantum-enhanced data processing scheme can be realized with fewer entangled encoded qubits~\cite{biamonte2017quantum, bergholm2018pennylane} to assure model parameters protection and lower computational complexity. As shown in Table~\ref{tab:overview}, to the best of the authors' knowledge, this is the \textbf{first} work to combine quantum circuits and DNNs and build a new QCNN~\cite{henderson2020quanvolutional} for ASR. To 
provide secure data pipeline and reliable quantum computing, we introduce the VFL architecture for decentralized ASR tasks, where remote NISQ cloud servers are used to generate quantum-based features, and ASR decoding is performed with a local model~\cite{yang2020multi}. We refer to our decentralized quantum-based ASR system to as QCNN-ASR. Evaluated on the Google Speech Commands dataset with machine noises incurred from quantum computers, the proposed QCNN-ASR framework attains a competitive 95.12\% accuracy on word recognition.
\begin{table}[t]\footnotesize
\centering
\vspace{-2mm}
\caption{An overview of machine learning  approaches and related key properties. CQ stands for a hybrid classical-quantum (CQ)~\cite{biamonte2017quantum} model using in this paper. QA stands for quantum advantages ~\cite{havlivcek2019supervised}, which are related to computational memory and parameter protection. VQC indicates the variational quantum circuit. VFL means vertical federated leaning~\cite{yang2019federated}. DNN stands for deep neural network~\cite{deng2013new}}
\label{tab:overview}
\begin{adjustbox}{width=0.48\textwidth}
\begin{tabular}{|l|l|l|l|l|}
\hline
Approach & Input & Learning Model & Output & Properties \\ \hline
Classical & bits & DNN and more. & bits & Easy implementation \\ \hline
Quantum & qubits & VQC and more. & qubits & QA but limited resources\\ \hline
hybrid CQ & bits & VQC + DNN & bits & Accessible QA over VFL \\ \hline
\end{tabular}
\end{adjustbox}
\vspace{-4mm}
\end{table}

\section{Related Work}
\label{sec:rw}
\subsection{Quantum Machine Learning for Signal Processing} 
QML~\cite{mitarai2018quantum} has been shown advantages in terms of lower memory storage, secured model parameters encryption, and good feature representation capabilities~\cite{havlivcek2019supervised}. There are several variants (e.g., adiabatic quantum computation~\cite{farhi2000quantum}, and quantum circuit learning~\cite{yao1993quantum}). In this work, we use the hybrid classical-quantum algorithm~\cite{henderson2020quanvolutional}, where the input signals are given in a purely classical format (aka, numerical format, e.g., digital image), and a quantum algorithm is employed in the feature learning phase. Quantum circuit learning is regarded as the most accessible and reproducible QML for signal processing~\cite{biamonte2017quantum}, such as supervised learning in the design of quantum support vector machine~\cite{havlivcek2019supervised}.
Indeed, it has been widely used, and it consists only of quantum logic gates with a possibility of deferring an error correction~\cite{mitarai2018quantum, yao1993quantum}. 

\subsection{Deep Learning with Variational Quantum Circuit}
In the NISQ era~\cite{preskill2018quantum}, quantum computing devices are not error-corrected, and they are therefore not fault-tolerant. Such a constraint limits the potential applications on NISQ technology, especially for large quantum circuit depth, and a large number of qubits. However,  Mitarai \emph{et al.}'s seminal work \cite{mitarai2018quantum} describes a framework to build machine learning models on NISQ. The key idea is to employ VQC~\cite{benedetti2019parameterized}, which are subject to an iterative optimization processes, so that the effects of noise in the NISQ devices can potentially be absorbed into these learned circuit parameters. Recent litterature reports about several successful machine learning applications based on VQC, for instance, deep reinforcement learning \cite{chen2020variational}, and function approximation  \cite{mitarai2018quantum}. VQCs are also used in constructing quantum machine learning models capable of handling sequential patterns, such as the dynamics of of certain physical systems \cite{chen2020quantum}. It should be noted that the input dimension of the input  in \cite{chen2020quantum} is rather limited~\cite{chen2020variational} because of stringent requirements of currently available quantum simulators, or real quantum devices.
\subsection{Quantum Learning and Decentralized Speech Processing}
Although quantum technology is quite new, there have been some attempts in exploiting it for speech processing. For example,  Li \emph{et al.}~\cite{li2002quantum} proposed a speech recognition system with quantum back-propagation (QBP) simulated by fuzzy logic computing. However, QBP is not using the qubit directly in a real-world quantum device, and the approaches hardly demonstrates the quantum advantages inherent in this computing scheme. Moreover, the QBP solution can be complicated to large-scale ASR tasks with parameters protection. 

From a system perspective, these accessible quantum advantages from VQL, including encryption and randomized encoding, are prominent requirements for federated learning systems, such as distributed ASR. Cloud computing-based federated architectures~\cite{yang2019federated} have been proven the most effective solutions for industrial applications, demonstrating quantum advantages using commercial NISQ servers~\cite{ mohseni2017commercialize}.  More recent works on federated keyword spotting~\cite{leroy2019federated}, distributed ASR~\cite{qi2020submodular}, improved lite audio-visual processing for local inference~\cite{chuang2020improved}, and federated n-gram language~\cite{chen2019federated} marked the the importance of privacy-preserving learning under the requirement of acoustic and language data protection.

\section{Designing Quantum Convolutional Neural Networks for Speech Recognition}
\label{sec:proposal}
In this section, we present our framework showing how to design a federated architecture based QCNN composed of quantum computing and deep learning for speech recognition.

\subsection{Speech Processing under Vertical Federated Learning}
We consider a federated learning scenario for speech processing, where the ASR system includes two blocks deployed between a local user, and a cloud server or application interface (API), as shown in Fig.~\ref{fig:1:vfl}. An input speech signal, $x_i$, is collected at the local user and up-streamed to a cloud server where Mel spectrogram feature vectors are extracted, $\textbf{u}_{i}$. Mel spectrogram features are the input of a quantum circuit layer, $\mathbf{Q}$, that learns and encodes patterns:
\begin{equation}
    f_{i}=\mathbf{Q}(\textbf{u}_{i}, \mathbf{e}, \mathbf{q}, \mathbf{d}),~~~~ \text{where}~~~\textbf{u}_{i} = \text{Mel-Spectrogram}(\textbf{x}_i).
    \label{eq:1:vfl:audio}
\end{equation}

In Eq. (\ref{eq:1:vfl:audio}), the computation process of a quantum neural layer, $\mathbf{Q}$, depends on the encoding initialization $\mathbf{e}$, the quantum circuit parameters, $\mathbf{q}$, and the decoding measurement $\mathbf{d}$. The encoded features, $f_{i}$, will be down-streamed back to the local user and used for training the ASR system, more specifically the acoustic model (AM). Proposed decentralized-VFL speech processing model reduces the risk of parameter leakages~\cite{duc2014unifying, leroy2019federated, chen2019federated} from attackers, and avoids privacy issues under GDPR, with its architecture-wise advantages~\cite{dwork2015reusable} on encryption~\cite{yao1993quantum} and without accessing the data directly~\cite{ yang2019federated}. 

\begin{figure}[ht!]
\begin{subfigure}[b]{0.200\textwidth}  
\centering 
\includegraphics[width=\textwidth]{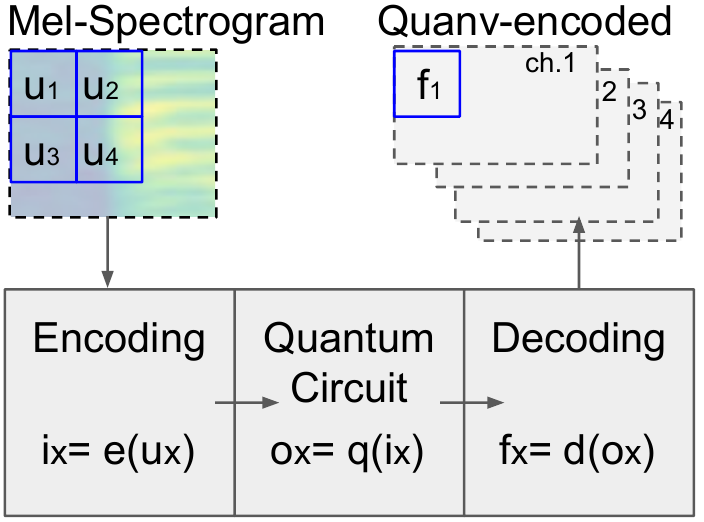}
\caption{QCNN Computing Process.}
\end{subfigure}
\quad
\begin{subfigure}[b]{0.250\textwidth}
\centering
\begin{minipage}{5cm}
\Qcircuit @C=0.20em @R=0.20em {
\lstick{\ket{0}} & \gate{R_y}  & \qw        & \gate{R_x}  & \qw      & \targ            &\qw & \gate{R_y} & \gate{R_x} & \meter \qw \\
\lstick{\ket{0}} & \gate{R_y}  & \qw        & \targ              & \qw      & \qw              &\qw &\qw                &\qw                & \meter \qw \\
\lstick{\ket{0}} & \gate{R_y}  & \qw        & \ctrl{-1}          & \qw      & \qw              &\qw &\qw                &\qw                & \meter \qw \\
\lstick{\ket{0}} & \gate{R_y}  & \qw        & \gate{R_z}  & \qw      & \ctrl{-3}        &\qw &\qw                &\qw                & \meter \qw \\
}
\end{minipage}
\caption{Deployed Quantum Circuit.}
\end{subfigure}
\caption{{\it The proposed variational quantum circuit for 2 $\times$ 2 QCNN.}}
\label{fig:circuit}
\end{figure}
\vspace{-2mm}

\subsection{Quantum Convolutional Layer}
Motivated by using VQC as a convolution filter with a quantum kernel, QCNN~\cite{henderson2020quanvolutional} was recently proposed to extend CNN's properties to the quantum domain for image processing on a digital simulator and requires only fewer qubits to construct a convolution kernel during the QML progress.
A QCNN consists of several quantum convolutional filters, and each quantum convolutional filter transforms input data using a quantum circuit that can be designed in a structured or a randomized fashion. 

Figure~\ref{fig:circuit} (a) show our implementation of a quantum convolutional layer. The quantum convolutional filter is consists of (i) the encoding function $e(\cdot)$, (ii) the decoding operation $d(\cdot)$, and (iii) the quantum circuit $q(\cdot)$. In detail, the following steps are performed to obtain the output of a quantum convolutional layer:
\begin{itemize}
\item The 2D Mel-spectrogram input vectors are chunked into several $2\times 2$ patches, and the $n^{th}$ patch is fed into the quantum circuit and encoded into intial quantum states, $\textbf{I}_{\textbf{x}}[n] = e(\textbf{u}_{i}[n])$. 
\item The initial quantum states go through the quantum circuit with the operator $q(\cdot)$, and generate $O_{x}[n] = q(\textbf{I}_{x}[n])$. 

\item The outputs after applying the quantum circuit are necessarily measured by projecting the qubits onto a set of quantum state basis that spans all of the possible quantum states and quantum operations. Thus we get the desired output value,  $f_{\textbf{x}, n} = d(\textbf{O}_{\textbf{x}}[n])$. More details refer to the implementation$^1$.
\end{itemize}

\vspace{-3mm}
\subsection{Random Quantum Circuit}
We deploy a random quantum circuit to realize a simple circuit $U$ in which the circuit design is randomly generated per QCNN model for parameter protection. An example of random quantum circuit is shown in Figure~\ref{fig:circuit} (b), where the quantum gates $R_{x}, R_{y}$ and $R_{z}$ and CNOT are applied. The classical vectors are initially encoded into a quantum state $\Phi_{0} = |0000\rangle$, and the encoded quantum states go through the quantum circuit $U$ for the following phases as:

\noindent \textbf{Phase 1}: $\Phi_{1} = R_{y}|0\rangle R_{y}|0\rangle R_{y} |0\rangle R_{y} |0\rangle$. 

\noindent \textbf{Phase 2}: $\Phi_{2} = (R_{x}R_{y}|0\rangle) \text{CNOT}(R_{y}|0\rangle)R_{y}|0\rangle R_{z}R_{y}|0\rangle$.

\noindent \textbf{Phase 3}: 
$\Phi_{3} = \text{CNOT}((R_{x}R_{y}|0\rangle)) \text{CNOT}(R_{y}|0\rangle)R_{y}|0\rangle R_{z}R_{y}|0\rangle.$

\noindent \textbf{Phase 4}:
$\Phi_{4} = R_{x}R_{y}\Phi_{3}$

Besides, since random quantum circuit may involve many CNOT gates which bring about many unexpected noisy signals under the current non error-corrected quantum devices and the connectivity of physical qubits, we limit the number of qubits to small numbers to avoid exceeding the noise tolerance capabilities of VQC. In the simulation on CPU, we use PennyLane~\cite{bergholm2018pennylane}, which is an open-source programming software for differentiable programming of quantum computers, to generate the random quantum circuit, and we build the random quantum circuit based on the Qiskit~\cite{aleksandrowicz2019qiskit} for simulation with the noise model from IBM  quantum machines with 5 and 15 qubits, which is advanced than simulation only results~\cite{henderson2020quanvolutional}.

\begin{figure}[h]
\begin{center}
  \centering    
\includegraphics[width=0.95\linewidth]{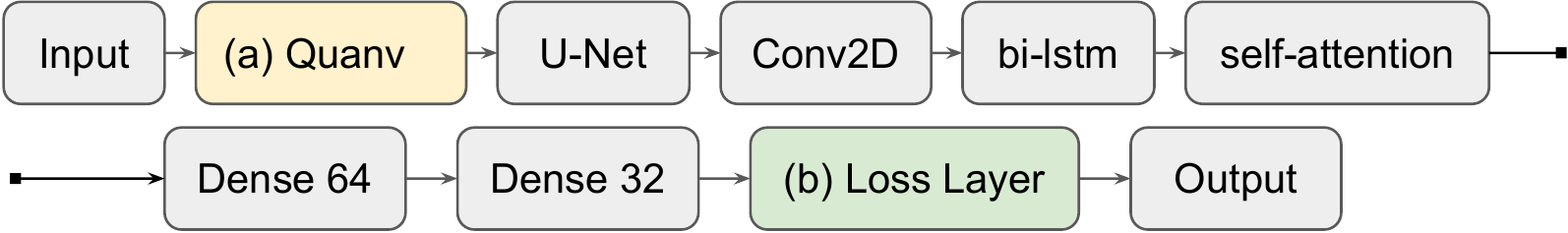}
\end{center}
\vspace{-0.4cm}
  \caption{The proposed QCNN architecture for ASR tasks.
  } 
\label{fig:3:network}
\vspace{-0.6cm}
\end{figure}

\subsection{Attention Recurrent Neural Networks}
We use a benchmark deep attention recurrent neural network (RNN)~\cite{hochreiter1997long} model from \cite{de2018neural} as our baseline architecture for a local model (e.g., second-pass models~\cite{chen2019federated,yang2020multi}) in the VFL setting. The model is composed of two layers of bi-directional long short-term memory~\cite{hochreiter1997long} and a self-attention encoder~\cite{vaswani2017attention}~(dubbed RNN$_\mathbf{Att}$). In~\cite{de2018neural}, this RNN model has been reported the best results over the other DNN based solutions included DS-CNN~\cite{zhang2017hello} and ResNet~\cite{warden2018speech} for spoken word recognition.

To reduce architecture-wise variants on our experiments, we conduct ablation studies and propose an advanced attention RNN model with a U-Net encoder~\cite{yang2020characterizing} (denoted as RNN$_\mathbf{UAtt}$). As shown in Fig.~\ref{fig:3:network}, a series of multi-scale convolution layers (with a channel size of 8-16-8) will apply on quantum-encoded (quanv) or neural convolution-encoded (conv) features to improve generalization of acoustic by learning scale-free representations~\cite{yang2020characterizing}. We use RNN$_\mathbf{Att}$ and RNN$_\mathbf{UAtt}$ in our experiments to evaluate the advantages of using the proposed QCNN model. As shown in Fig~\ref{fig:3:network}~(b), we provide a loss calculation layer on the RNN backbone for our local model. For spoken word recognition, we use the cross-entropy loss for classification. The loss layer could further be replaced by connectionist temporal classification (CTC) loss~\cite{graves2006connectionist} for a large-scale continuous speech recognition task in our future study.

\section{Experiments}
\label{sec:exp}
\subsection{Experimental Setup}
As initial assessment of the  viability our novel proposed framework, we have selected a limited-vocabulary yet reasonably challenging speech recognition task, namely the Google Speech Command-V1~\cite{warden2018speech}. For spoken word recognition, we use the ten-classes setting that includes the following frequent speech commands\footnote{ \href{https://ai.googleblog.com/2017/08/launching-speech-commands-dataset.html}{https://ai.googleblog.com/2017/08/launching-speech-commands-dataset.html}}: ['left', 'go', 'yes', 'down', 'up', 'on', 'right', 'no', 'off', 'stop'], with a total of 11,165 training examples, and 6,500 testing examples with the background white noise setup~\cite{warden2018speech}. 
The Mel-scale spectrogram features are extracted from the input speech using the Librosa library; this step takes place in the NISQ server. The input Mel-scale feature is actually a 60-band Mel-scale, and 1024 discrete Fourier transform points into the quantum circuit as the required VFL setting. The experiments with the local model are carried out with Tensorflow, which is used  to implement DNNs and visualization.

\begin{figure}[ht!]
\begin{center}
\vspace{-2mm}
  \centering    
\includegraphics[width=0.85\linewidth]{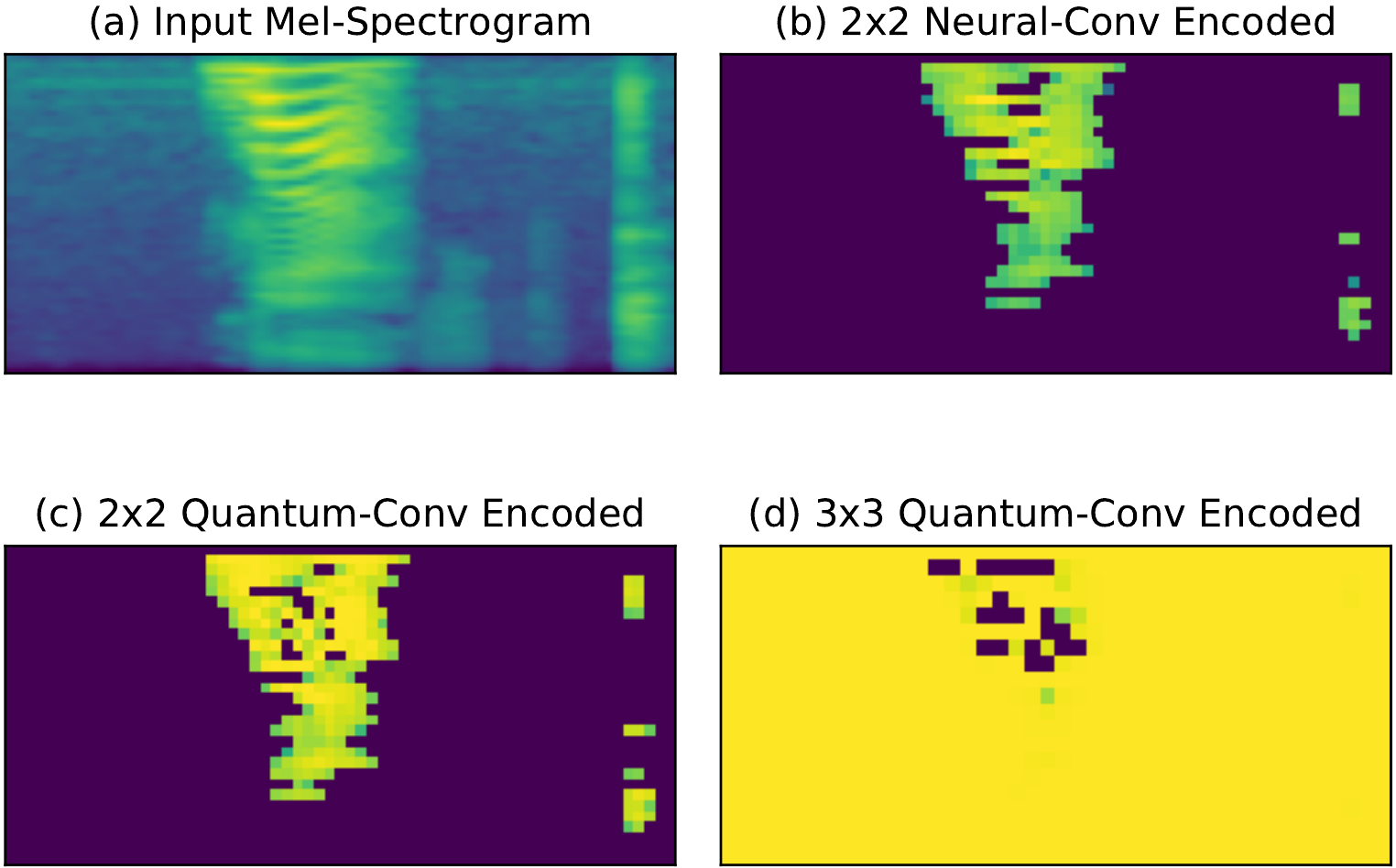}
\end{center}
\vspace{-0.2cm}
  \caption{Visualization of the encoded features from different types of convolution layers. The audio transcription is "yes" of the input.
  } 
\label{fig:2:enc}
\end{figure}
\vspace{-0.4cm}

\subsection{Encoded Acoustic Features from Quantum Device}
The IBM Qiskit quantum computing tool~\cite{aleksandrowicz2019qiskit} is used to simulate the quantum convolution. 
We first use Qiskit to collect compiling noises from two different quantum computers. We then load those recorded noise to the Pennylane-Qiskit extension in order to simulate noisy quantum circuit experiments for virtualization. According to previous investigations~\cite{chen2020quantum, chen2020variational}, the proposed noisy quantum device setup can be complied with  NISQ directly and attains results close to those obtained using NISQ directly
. The chosen setup preserves quantum advantages on randomization and parameter isolation. 

\textbf{Visualization of Acoustic Features.} To better understand the nature of the encoded representation of our acoustic speech features, we visualize the encoded features and acoustic patterns extracted from different encoders. Fig.~\ref{fig:2:enc} shows QCNN-encoded results with a 2$\times$2 kernel (in panel (c)), which seems to better relate to the acoustic pattern shown in the Mel spectrogram shown in panel (a), since it well captures energy patterns in both high and low-frequency regions. The latter becomes more evident by comparing panel (c) with the features encoded with a 3$\times$3 kernel given in panel (d). Finally, the neural network-based convolution layer reported in panel (b) shows similar results with those in panel (c), but it presents a lower intensity in the high-frequency  regions. We will discuss its relationship between recognition performance later in Section \ref{sec:4:4}.

\subsection{Performance of Spoken-Word Recognition}
We conduct experiments on the spoken-word recognition task and compared the improved performance from an additional quantum convolution layer with a 2$\times$2 kernel (4 qubits) and a neural convolution layer with a 2$\times$2 kernel in Table~\ref{tab:1:acc}. From the experiments, the recognition models with additional quantum convolution show better accuracy than the baseline models \cite{de2018neural}. The modified model with a U-Net encoder, RNN$_\mathbf{UAtt}$, achieves the best performance 
of 95.12$\pm$0.18\% on the evaluation data, which is better than 
the reproduced RNN$_\mathbf{Att}$ baseline (94.21$\pm$0.30\%) for the recognition setup.

\begin{table}[ht!]
\centering
\caption{Comparisons of spoken-term recognition on Google Commands dataset with the noise setting~\cite{warden2018speech} for classification accuracy (Acc) $\pm$ standard deviation. The additional convolution (conv) and quantum convolution (quanv) layer have the same 2$\times$2 kernel size.}
\label{tab:1:acc}
\begin{adjustbox}{width=0.48\textwidth}
\begin{tabular}{|l|c|l|}
\hline
Model & Acc. ($\uparrow$) & Parameters (Memory) ($\downarrow$) \\ \hline
RNN$_\mathbf{Att}$~\cite{de2018neural} & 94.21$\pm$0.30 & 170,915 (32-bits) \\
Conv + RNN$_\mathbf{Att}$& 94.32$\pm$0.26 & 174,975 (32-bits) \\
Quanv + RNN$_\mathbf{Att}$ & 94.75$\pm$0.17 & 174,955 (32-bits) + 4 (qubits) \\ \hline
RNN$_\mathbf{UAtt}$ & 94.72$\pm$0.23 & 176,535 (32-bits) \\
Conv + RNN$_\mathbf{UAtt}$ & 94.74$\pm$0.25 & 180,595 (32-bits) \\
Quanv + RNN$_\mathbf{UAtt}$ & \textbf{95.12}$\pm$0.18 & 180,575 (32-bits) + 4 (qubits) \\ \hline
\end{tabular}
\end{adjustbox}
\vspace{-4mm}
\end{table}

\subsection{A Study on QCNN Architectures}
\label{sec:4:4}
\begin{figure}[ht!]
\begin{center}
\vspace{-2mm}
  \centering    
\includegraphics[width=0.85\linewidth]{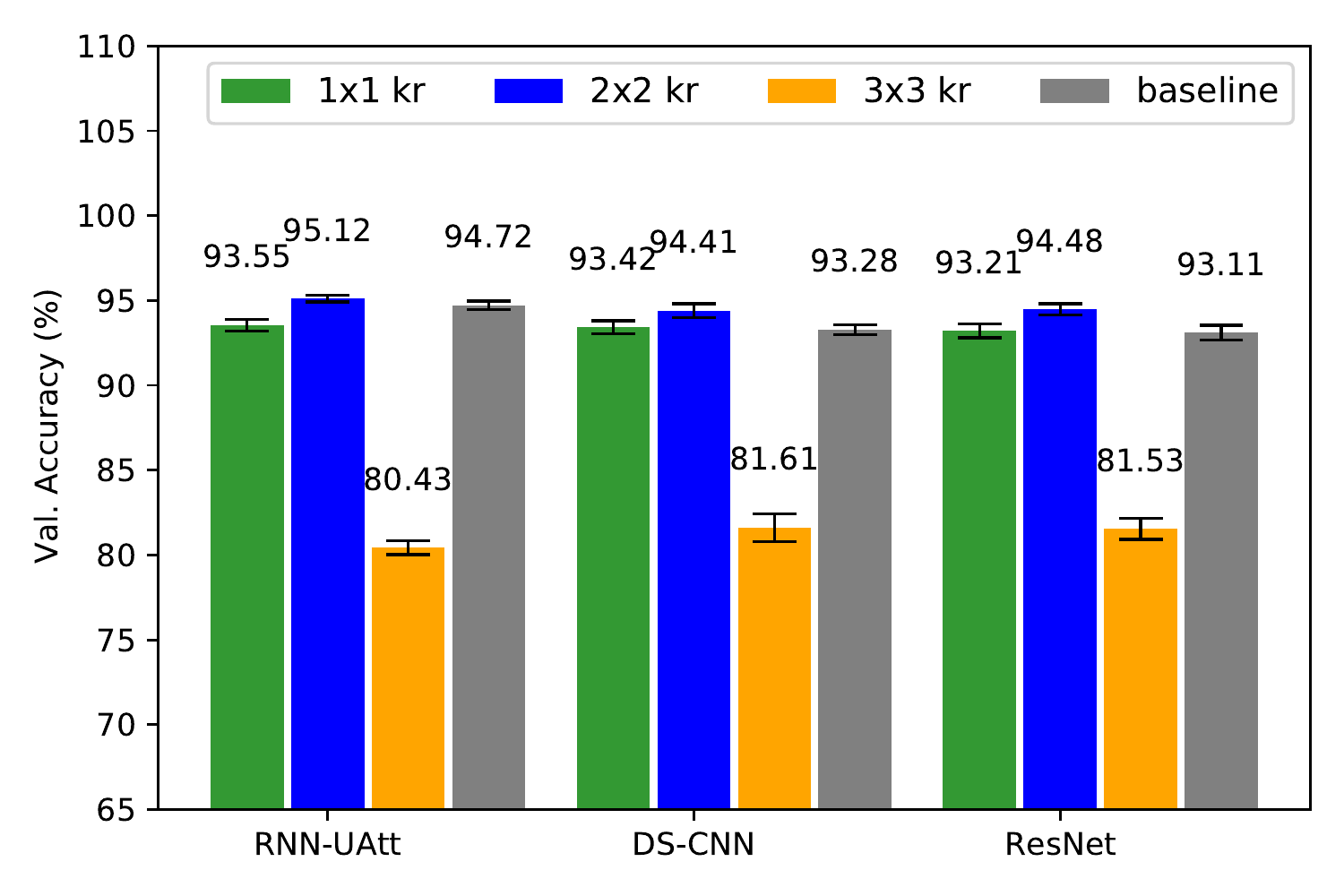}
\end{center}
\vspace{-0.4cm}
  \caption{Performance studies of different quantum kernel size (dubbed kr) with DNN acoustic models for designing QCNN models. 
  } 
\label{fig:aba}
\end{figure}

Next we experiment with various new QCNN~\cite{henderson2020quanvolutional} architectures for ASR with different combinations of quantum encoders and neural acoustic models. First, we study the quantum convolution encoder with different kernel sizes. From previous works~\cite{chen2020quantum, chen2020variational}, the current commercial NISQ devices would be challenging to provide reproducible and stable results with a size of qubits larger than 15.
We thus design our quantum convolutional encoders under this limitation with a kernel size of 1$\times$1 (1 qubit), 2$\times$2 (4 qubits), and 3$\times$3 (9 qubits).
We select two open source neural AMs as the local model, DS-CNN~\cite{zhang2017hello}, and ResNet~\cite{warden2018speech}, from the previous works testing on the Google Speech Commands dataset. As shown in the bar charts in Fig.~\ref{fig:aba}, QCNNs with the 2$\times$2 kernel show better accuracy and lower deviations than all other models tested. QCNN attains 1.21\% and 1.47\% relative improvements over DS-CNN and ResNet baseline, respectively. On the other hand, QCNNs with the 3$\times$3 kernel show the worst accuracy when compared with other configurations.
Increasing the kernel size does not always guarantee improved performances in the design of QCNN for the evaluation. The encoded features obtained with a 3$\times$3 quantum kernel used to train AMs, as shown in Fig.~\ref{fig:2:enc}(d), are often too sparse and not as discriminative when compared to those obtained with 1$\times$1 and 2$\times$2 quantum kernels, as indicated in Fig.~\ref{fig:2:enc}(b) and Fig.~\ref{fig:2:enc}(c), respectively.

\begin{figure}[ht!]
\begin{center}
\vspace{-2mm}
  \centering    

\includegraphics[width=0.80\linewidth]{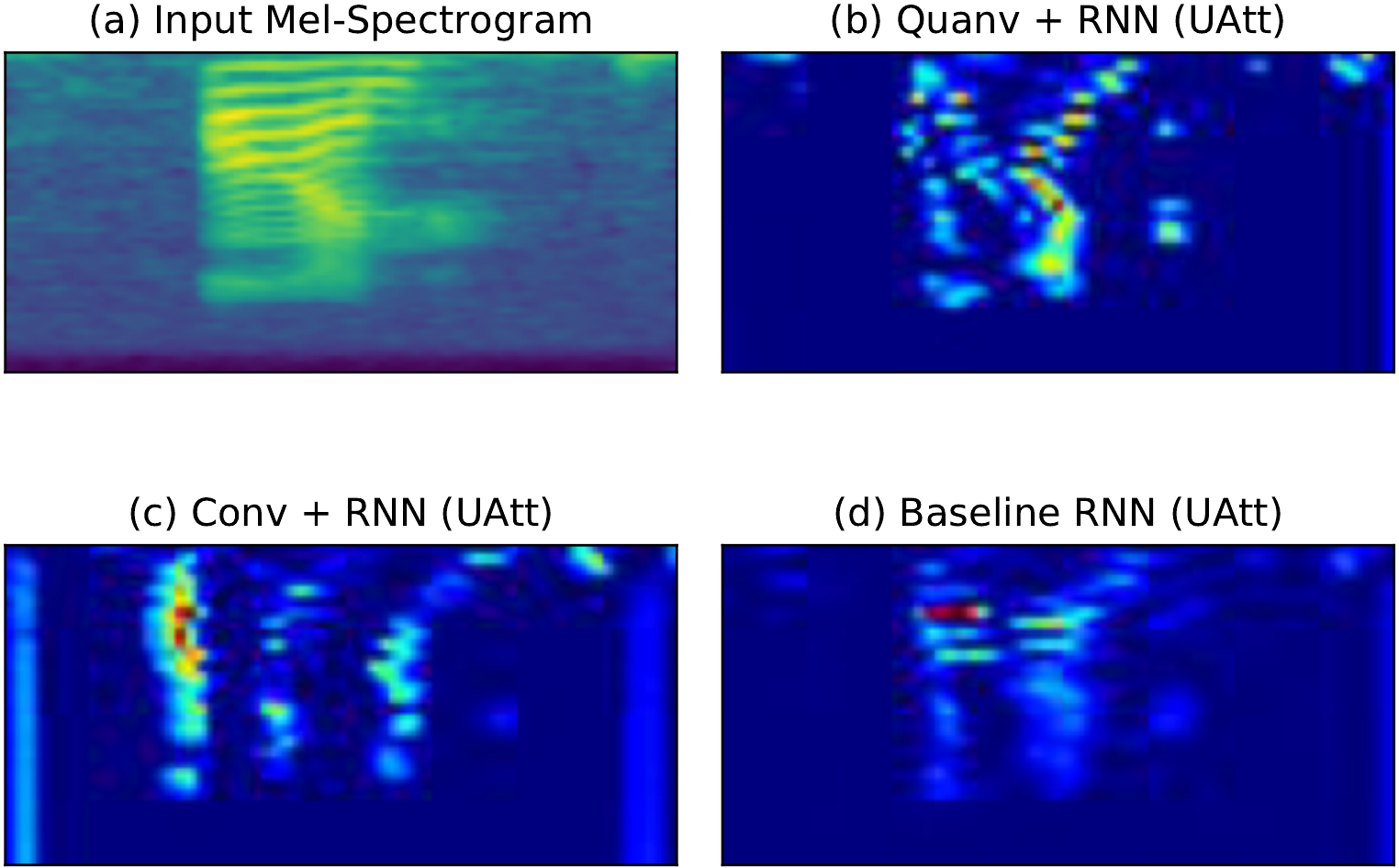}
\end{center}
\vspace{-0.2cm}
  \caption{Interpretable neural saliency results by class activation mapping~\cite{zhou2016learning} over (a) Mel spectrogram features with audio transcription of "on"; (b) a 2$\times$2 quantum convolution layer followed by RNN$_\mathbf{UAtt}$; (c) a well-trained 2$\times$2 neural convolution layer followed by RNN$_\mathbf{UAtt}$, and (d) baseline RNN$_\mathbf{UAtt}$.
  } 
\label{fig:cam}
\end{figure}
\vspace{-0.2cm}

\subsection{A Saliency Study by Acoustic Class Activation Mapping}
We use a benchmark neural saliency technique by class activation mapping (CAM)~\cite{zhou2016learning} over different neural acoustic models to highlight the responding weighted features that activate the current output prediction. As shown in Fig.~\ref{fig:cam}, QCNN (b) learns much more correlated and richer acoustic features than RNN with a convolution layer and baseline model~\cite{de2018neural}. According to the CAM displays, the activated hidden neurons learn to identify related low-frequency patterns when making the ASR prediction from an utterance "on."

\section{Conclusion}
\label{sec:conclusion}
In this paper, we propose a new feature extraction approach to decentralized speech processing to be used in vertical federated learning that facilitates model parameter protection and preserves interpretable acoustic feature learning via \textbf{quantum convolution}. The proposed QCNN models show competitive recognition results for spoken-term recognition with stable performance from quantum machines when learning compared with classical DNN based AM models with the same convolutional kernel size. Our future work includes incorporating QCNN into continuous ASR. Although the proposed VFL based ASR architecture fulfilling some data protection requirements by decentralizing prediction models, more statistical privacy measurements~\cite{dwork2015reusable} will be deployed to enhance the proposed QCNN models from the other privacy perspectives~\cite{dwork2015reusable, leroy2019federated}.

\clearpage
\bibliographystyle{IEEEtran}

\bibliography{refs,qml}

\end{document}